This is a preprint of :
Morrison G.S., Stoel R.D. (2014). Forensic strength of evidence statements should preferably be likelihood ratios calculated using relevant data, quantitative measurements, and statistical models – a response to Lennard (2013) Fingerprint identification: How far have we come? *Australian Journal of Forensic Sciences*, 46, 282–292. http://dx.doi.org/10.1080/00450618.2013.833648

# Forensic Strength of Evidence Statements Should Preferably be Likelihood Ratios Calculated Using Relevant Data, Quantitative Measurements, and Statistical Models – A Response to Lennard (2013) Fingerprint Identification: How Far Have We Come?

Geoffrey Stewart Morrison[a*] and Reinoud D Stoel[b]

[a]*Forensic Voice Comparison Laboratory, School of Electrical Engineering & Telecommunications, University of New South Wales, Sydney, Australia*

[b]*Netherlands Forensic Institute, The Hague, The Netherlands*

Lennard (2013) [Fingerprint identification: how far have we come? Aus J Forensic Sci. doi:10.1080/00450618.2012.752037] proposes that the numeric output of statistical models should not be presented in court (except "if necessary" / "if required"). Instead he argues in favour of an "expert opinion" which may be informed by a statistical model but which is not itself the output of a statistical model. We argue that his proposed procedure lacks the transparency, the ease of testing of validity and reliability, and the relative robustness to cognitive bias that are the strengths of a likelihood-ratio approach based on relevant data, quantitative measurements, and statistical models, and that the latter is therefore preferable.

This research was supported by the Australian Research Council, the Australian Federal Police, New South Wales Police, Queensland Police, the National Institute of Forensic Science, the Australasian Speech Science and Technology Association, and the Guardia Civil via Linkage Project LP100200142. Unless otherwise explicitly attributed, the opinions expressed herein are those of the authors and do not necessarily represent the policies or opinions of any of the above mentioned organisations. Thanks to David Balding and to two anonymous reviewers for comments on earlier versions of this paper.

## Introduction

Professor Chris Lennard was a Plenary Invited Speaker at the Australia New Zealand Forensic Science Society's 21st International Symposium on the Forensic Sciences, Hobart, September 2012. A paper based on his presentation was subsequently published in the Australian Journal of Forensic Sciences.[1] Lennard argues that the traditional "inconclusive" outcome in forensic fingerprint analysis should be replaced by "qualified conclusions backed up by an appropriate statistical model" (Ref 1, p. 8 [page references are to the version of record first published 18 Jan 2013, i.e., p. 8 is the eighth page of the article]). He contrasts current practice in forensic fingerprint analysis, where conclusions have traditionally been expressed as "individualization" v "inconclusive" v "exclusion", with current practice in forensic DNA analysis, where conclusions are expressed as numeric likelihood ratios calculated using statistical models. (In simple source-level DNA cases, for which it is assumed that the data are discrete and have no source or measurement level variability, random match probabilities are often presented; however, a random match probability is simply the denominator of the likelihood ratio and its use implies a numerator of one; Ref 2, p. 404; Ref 3; Ref 4, §99.190; Ref 5, p. 22–23.)

*Corresponding author. Now: Dr Geoffrey Stewart Morrison Forensic Consultant, Vancouver, British Columbia, Canada. e-mail: geoff-morrison@forensic-evaluation.net

> Fingerprints and DNA profiling are therefore at opposite ends of a spectrum ... Categorical opinions with no supporting statistical model are at one end, while at the other we have a nearly exclusive reliance on statistics and probabilities, with a general reluctance to provide an actual opinion. (Ref 1, p. 9)

We agree with Lennard that the “inconclusive” category traditionally used in forensic fingerprint analysis denies the trier of fact potentially useful information and that it should be replaced by “qualified conclusions” which indicate the strength of the evidence. We also agree that statistical models should be used. We disagree, however, with other aspects of Lennard’s proposal, and in this paper present what we believe is a preferable alternative.

## An expert opinion informed by, but which is not, the output of a statistical model

Lennard proposes a “middle ground” between existing practice in fingerprints and DNA whereby the forensic scientist expresses an “expert opinion” which is “informed, supported and backed-up by an appropriate statistical model” (Ref 1, p. 9) but which is not itself the direct output of the statistical model:

> [T]he LR [likelihood ratio] is a calculated result from a mathematical model, it is not an opinion. (Ref 1, p. 9)
>
> The evidence is the expert opinion and not the statistic. (Ref 1, p. 9)
>
> The statistics should arguably only be presented if required. (Ref 1, p. 9)

He proposes that an “expert opinion” rather than the numeric likelihood-ratio output of a statistical model be used for the expression of the strength of DNA evidence as well as for the strength of fingerprint evidence (and by implication also for the strength of evidence from other branches of forensic science). Lennard does not actually define what he means by “expert opinion” or the process by which it is informed by a statistical model, but we take it to mean a judgement as to the strength of the evidence made by the expert on the basis of their experience and training, a subjective decision which takes the output of the statistical model into account but which also takes into account other information.

The one concrete example Lennard provides of an interaction between an expert and a statistical model is actually the opposite of his main proposal, and can be used to illustrate the problems one would encounter in trying to implement his approach. Rather than a statistical model calculating a likelihood ratio and then the expert considering that in reaching their “expert opinion”, the example is of a fingerprint expert making an “individualization” conclusion and ascribing observed differences to skin distortion, whereas the results from the model subsequently indicated that these differences “cannot be explained by normal skin distortion” (Ref 1, p. 6). If the skin-distortion model were part of a larger model which calculated a likelihood ratio, and that likelihood ratio indicated that the evidence was more probable given the different-origin hypothesis than the same-origin hypothesis, but the expert's initial opinion using traditional methods was "individualization", then how would the disagreement between the model and traditional approach be resolved in the final "expert opinion" informed by both sources of information? Would the traditionally derived result effectively overrule the result based on relevant data, quantitative measurements, and statistical models, or vice versa? If the model had provided a relatively low-magnitude likelihood ratio in favour of the same-origin hypothesis, there would still be a disagreement between the model and the traditional approach – a simple rule, such as go with the traditionally-derived result if the output of the statistical model points in the same direction, would not be adequate. We would argue in favour of having a predefined explicit rule for combining multiple sources of information to arrive at a strength-of-evidence statement, and that such a rule be empirically derived from relevant data (for example, using logistic-regression fusion[6]) rather than simply being a heuristic.

For forensic DNA analysis (and other branches of forensic science such as forensic speech science) we consider the proposal to present an "expert opinion" rather than the output of the statistical model a step backwards. It may represent a step forwards for fingerprint analysis but it should not be the final destination. The great advantages of procedures based on relevant data, quantitative measurements, and statistical models are that they are transparent and relatively easily tested. By transparent we mean that each step can be described in sufficient detail that the entire procedure can be replicated by any suitably qualified individual in any suitably equipped laboratory. The data and software employed can even be provided to facilitate the task. By relatively easily tested we mean that it is relatively easy to run a relatively large set of appropriate test data through the system and obtain an assessment of the validity and reliability of the system's performance.[7] The forensic scientist's expertise is still required to select appropriate models, appropriate training data, and appropriate test data (see below), but, beyond that, the greater the automation of the system the more easily it can be tested.

In contrast, if the ultimate decision as to the strength of the evidence is made subjectively by a human expert, this lacks transparency. We do not have direct access to the forensic expert's cognitive processes: their description of what they consciously thought they did may not be what they actually did. Judgements made in this way are also more susceptible to cognitive bias due to the influence of domain-irrelevant information,[8,9] which was of great concern in the 2009 National Research Council (NRC) report.[10]

Demonstrating validity and reliability was also a major concern of the 2009 NRC report, and of the 2012 report by the National Institute of Standards and Technology (NIST) / National Institute of Justice (NIJ) Expert Working Group on Human Factors in Latent Fingerprint Analysis,[11] and calls for the validity and reliability of approaches to forensic analysis to be tested under realistic casework conditions go back to at least the 1960s.[12] It may be relatively expensive to build a system based on relevant data, quantitative measurements, and statistical models, but once the system has been trained, and appropriate test data have been obtained, the cost (including time) to run each test pair is low. Hundreds or thousands of test pairs can be analysed at minimal cost. Likewise, it may be expensive to train a human expert, but, in contrast, when it comes to testing the assessment of each test pair by a human making a subjective experience-based judgement takes a relatively long time, and running a large number of test pairs is therefore costly. Thus it is harder, in a practical sense, to test the validity and reliability of experts whose ultimate decision is a subjective experience-based judgement. (Some systems based on quantitative measurements make measurements in a fully automatic way, others rely on human-supervised measurement procedures, see for example Refs 13 and 14. The cost of human-supervised measurements should be taken into account where relevant. If the "expert opinion" is actually informed by the statistical model, then the "expert opinion" will require more work than the statistical model alone, and will therefore always be more costly.) More importantly, if a system based on a relevant database, quantitative measurements, and statistical models, and an "expert opinion" system were both tested on the same test data reflecting the relevant population and the relevant conditions for the case under investigation, and the "expert opinion" outperformed the statistical-model system, then we would recommend the use of the "expert opinion" system in court in preference to the statistical-model system. Vice versa if the statistical-model system outperformed the "expert opinion" system.

Lennard states reasons why he does not consider it appropriate to present the results of a statistical analysis and why he considers an "expert opinion" to be more appropriate:

> If you change the assumptions that underpin your statistical model then the number will change, which is often the issue argued in court with respect to DNA evidence. However, the expert opinion should remain the same. The statistic, in any case, is only an approximation because such mathematical models are never perfect; a number of underlying assumptions are required for them to operate. (Ref 1, p. 9)

Lennard's arguments do not hold up as valid arguments against the direct use of the output of statistical models. Most of what he says are statements of fact which are common knowledge to anyone with a basic understanding of statistics. What he seems to imply is that there are problems with statistical models which would not affect an "expert opinion", when in fact an "expert opinion" would be subject to the same broad constraints as apply to statistical models. Further, the existence and nature of these constraints would be less transparent and explicit in an "expert opinion" than if the necessary steps for calculating a likelihood ratio using a statistical model were explained.

Lennard objects that statistical models are based on assumptions. It is true that statistical models are based on assumptions. No statistician would state otherwise. "[A]ll knowledge is a result of theory – we buy information with assumptions" (Ref 15, p. 5). When a statistical model is used the assumptions are made absolutely explicit by the choice of model. At a very basic level, the use of a Gaussian model, for example, implies the assumption that the population can reasonably be modelled using a Gaussian distribution and that the sample data are sufficiently representative of the population for the Gaussian model to provide useful information about the population. Contrary to what Lennard implies, however, an "expert opinion" must also be based on assumptions, and these assumptions are seldom (or even cannot be) made explicit in the same way as they are by the use of a statistical model.

Lennard objects that a statistic is an approximation. That a statistic is an approximation is true, it is an approximation of a population parameter. This is a fundamental concept in statistics, for example, a popular introductory statistics textbook states:

> As we begin to use sample data to draw conclusions about a wider population, we must take care to keep straight whether a number describes a sample or a population. Here is the vocabulary we use.
>
> PARAMETER, STATISTIC
>
> A **parameter** is a number that describes the population. In statistical practice, the value of a parameter is not known because we cannot examine the entire population.
>
> A **statistic** is a number that can be computed from the sample data without making use of any unknown parameters. In practice, we often use a statistic to estimate an unknown parameter. (Ref 16, p. 292)

Likewise, a statistical model is an approximation of reality; however:

> The fact that [a model] is an approximation does not necessarily detract from its usefulness because all models are approximations. Essentially, all models are wrong, but some are useful. (Ref 17, p.424)

The important practical issue is whether the model is sufficiently useful, i.e., whether it performs well enough given the conditions of the case under investigation. There are procedures to test the performance (the validity and reliability) of a forensic-comparison system under conditions reflecting those of the case under investigation using data drawn from the relevant population.[7] These are equally applicable to statistical-model and "expert opinion" systems. Contrary to what Lennard implies, an "expert opinion" must also be an approximation, and it is difficult to see how it could be any less of an approximation than a statistical model.

Lennard objects that if the assumptions underlying a statistical model change, then the results of the statistical analysis will change, but claims that the expert opinion should not change. It is only logical that if the assumptions underlying a statistical model change then the statistical model will change, and the output of a statistical analysis using the model is likely to change. Further, if an "expert opinion" is informed by the results of a statistical analysis, and those results change in a meaningful way, then surely the "expert opinion" should also change. If not, then in what sense is the expert opinion informed by the statistical model?

One of the anonymous reviewers of our paper stated some of the same arguments as we have made immediately above as:

> All decisions, whether statistically informed [or not], are conditional on a set of assumptions and background information. To criticise likelihood ratios for having them is absurd. Equally, if your opinion does not change when a radical deviation from the assumptions is observed (or made), then what is the value of that opinion? Either the assumption wasn't necessary, or the opinion is worthless.

## What goes into calculating a likelihood ratio and what should be fodder for debate before the trier of fact

> If you change the assumptions that underpin your statistical model then the number will change, which is often the issue argued in court with respect to DNA evidence. (Ref 1, p. 9)

> [Presentation in court of a number derived from a statistical model leads to] attacks on how the number was derived and what it actually means in relation to the case at hand. (Ref 1, p. 11)

This is in fact exactly the way we believe things should be. Let us interpret "assumptions" in a broad sense and consider what assumptions are required for a source-level forensic comparison. We describe the process assuming a procedure which makes use of relevant data, quantitative measurements, and statistical models. Homologous assumptions would be needed for an experience-based "expert opinion". Much of the following is discussed in greater depth in the context of forensic voice comparison in Ref 18.

First, the forensic scientist needs to define and communicate the prosecution and defence hypotheses as they understand them. A forensic likelihood ratio is the answer to a specific question, and to make sense of the likelihood ratio both the forensic scientist and the trier of fact need to understand that question. The question is specified by two hypotheses: the prosecution hypothesis, which specifies the numerator of the likelihood ratio, and the defence hypothesis, which specifies the denominator. In a source-level forensic comparison, the defence hypothesis specifies the relevant population, which is the hypothesised source of the sample of questioned origin. The relevant population is specific to the particular case under investigation (see, for example, Ref 19 on glass and Ref 20 on firearms). An entirely legitimate issue to debate before the trier of fact would be the question of whether the forensic scientist set out to test appropriate hypotheses, including whether the correct relevant population has been specified. By making their adopted hypotheses explicit, the forensic scientist facilitates consideration of this important question.

Next, the forensic scientist must obtain a sample from the relevant population. This sample is to be used to train the model which will calculate the denominator of the likelihood ratio. A legitimate issue to debate before the trier of fact would be whether the sample is sufficiently representative of the relevant population.[5,21]

Next, the forensic scientist must choose the statistical models that they will use to calculate the likelihood ratio. Every statistical model makes assumptions about the distribution of the data, but these assumptions are made explicit by the choice of model. Every model is an approximation. Some models make fewer assumptions about the distribution of the data and theoretically may be able to more closely approximate the true distribution for the population; however, they are trained on a limited amount of sample data, and models which make fewer assumptions generally require more sample data to adequately train them. A problem which emerges is that with limited training data such models may adapt to the peculiarities of the particular sample used for training (a problem known as overfitting). These models will then not perform well (will not generalise well) when making predictions about previously unseen data from the relevant population, such as the suspect and offender samples. A model which makes more assumptions about the distribution of the data may in theory not be able to approximate the true distribution for the population as closely. Such a model, however, will generally be less prone to overadapting to the peculiarities of the sample used for training, and may therefore perform better on previously unseen data. This phenomenon is known as bias-variance tradeoff.[17,22] Part of the expertise of the forensic scientist is to select a model which gives a reasonable approximation of the likely distribution of the population without overfitting the model to the training data. They can conduct tests using development data to help them select a model which gives what they consider to be sufficiently good performance.

The models should also be trained and optimised using data which reflect the conditions of the case under investigation so that the models take account of the conditions of the case under investigation. The specification of these conditions also forms part of the specific question which is to be answered by the likelihood ratio. In some branches of forensic science, the conditions of the suspect and offender samples will usually differ, e.g., a partial latent fingermark with distortion recovered from a crime scene, versus a carefully rolled or scanned fingerprint from a suspect. The question becomes: What is the probability of getting the properties of the distorted partial latent mark if it were produced by the same finger as made the high-quality suspect fingerprint versus if it were made by a finger of someone else from the relevant population? Such a model should not be trained and optimised on high-quality versus high-quality prints, or on distorted partial latent marks versus distorted partial latent marks, because neither would be answering the question about a distorted partial latent mark versus high-quality prints. See Ref 14 for examples of the effects of different recording conditions on the performance of forensic-voice-comparison systems. At a minimum, forensic-comparison systems should be trained under conditions reflecting those of the case under investigation, and ideally they should incorporate elements which explicitly attempt to compensate for mismatches between suspect and offender samples.

Once relevant training data have been selected and a model has been chosen, trained, and optimised to the conditions of the case under investigation, the system should be frozen, i.e., no other changes are allowed. Then the system should be tested using new pairs of samples drawn from the relevant population and reflecting the conditions of the actual suspect and offender samples from the case under investigation. In this way the forensic scientist obtains an indication of how well the system is expected to perform on previously unseen data from the relevant population and under these conditions, i.e., how well it is expected to perform on the actual suspect and offender samples. Testing using samples from some other population or under different conditions will not provide particularly useful information about how well the system is expected to perform on the actual suspect and offender samples. An issue for debate could be whether the conditions of the training and test data adequately reflect the conditions of the suspect and offender samples. If the judge at an admissibility hearing or the trier of fact at trial is satisfied that the samples adequately reflect the relevant population and conditions, and is satisfied that the model is answering the correct question, then the questions of whether the model employed is based on appropriate assumptions and whether it adequately approximates reality are replaced by the more concrete question of whether the model performs sufficiently well given this population and these conditions. What matters is whether the empirically demonstrated degree of validity and reliability of the model given the relevant population and conditions is sufficient for the output to be useful to the trier of fact in this case. It therefore becomes essential to present the results of validity and reliability tests to the judge at an admissibility hearing or to the trier of fact during trial. In Ref 7, the first author of the current paper argues that it is particularly important to present the results of reliability testing to the trier of fact.

At the risk of eisegesis, it seems to us that one of Lennard's underlying concerns may be that a single numeric likelihood ratio could be interpreted as a highly precise objective number, when it is in fact based on a number of subjective decisions (or "assumptions") related to selecting relevant data and training statistical models. If we may invert his sentence "The temptation is therefore to hide behind a number ... rather than give an actual opinion" (Ref 1, p. 9), Lennard's proposed solution appears to be to hide behind an "expert opinion" rather than give an actual statistical result. Our preferred solution is to make clear what debatable decisions actually need to be made as part of the process of calculating a likelihood ratio, and to actually assess the degree of reliability (precision) of the system that makes the calculations (see, for example, Refs 7, 23, 24, 25, 26). We think Lennard's solution hides all the details of how to calculate a likelihood ratio, and hides all the potential points for legitimate debate. In contrast, the presentation of the result of a statistical analysis and the explanation of what went into obtaining that result brings all of these details and potential points for debate out into the open. We would argue that the approach we advocate promotes full disclosure, whereas Lennard's recommendation is contrary to the interests of full disclosure.

## Definitive statements of individualisation or exclusion

Another idea with which Lennard seems to agree, but with which we disagree, is that forensic scientists should be permitted to make definitive statements of individualisation or exclusion on the basis of the statistical model having calculated a large likelihood ratio in one direction or the other:

> A definitive identification conclusion may still be the outcome if the LR is justifiably large enough to warrant such a decision. It would still be possible, therefore, to support the categorical 'identifications' and 'exclusions' – the extreme ends of the opinion scale – when the LRs are of a significant magnitude. (Ref 1, p. 8)

> There is a heavy reliance on the number that is generated; rarely is the evidence an actual opinion of uniqueness expressed by the forensic biologist. (Ref 1, p. 2)

> At what point do we accept identity based on DNA? Will biologists ever have the courage to say 'in my opinion it is his DNA' and will the courts allow this? (Ref 1, p. 10)

In the likelihood-ratio framework, expressing an opinion as to a posterior probability, the probability of the hypothesis given the evidence, is not compatible with the rôle of the forensic scientist. It is not logically possible for the forensic scientist to express an opinion as to a posterior probability without taking into account the prior probability, and this would require them to consider information extraneous to that required for an evaluation of the strength of the particular piece of evidence they have been asked to analyse. That the forensic scientist cannot express a posterior probability is a fundamental tenet of the likelihood-ratio framework.[27] Not only is Lennard's recommendation a recommendation to present posterior probabilities, but to present posterior probabilities of zero or one. A posterior probability of zero or one can only be derived in two ways, either the prior odds are zero or infinite, or the likelihood ratio is zero or infinite. The prior odds are outwith the competency of the forensic scientist. If the forensic scientist presents a likelihood ratio of zero or infinity then they are claiming that the prior odds are irrelevant and all other evidence relating to the question the likelihood ratio answers is irrelevant – no other evidence, no matter how strong, can outweigh this piece of evidence because the answer it provides is a definitive exclusion or a definitive individualisation. In making such a statement the forensic scientist is usurping the rôle of the trier of fact. The forensic scientist should inform the trier of fact about the strength of the evidence, not make the same-origin or different-origin decision for the trier of fact. (A potential exception would be a simple source-level DNA case in which the discrete values of different profiles differ and the numerator of the likelihood ratio is determined to be zero – an exclusion. The trier of fact could still consider the probability of measurement error, clerical error, contamination, etc. For continuously valued variables with variability at the source, measurement, or modelling level, such as the acoustic properties of voice recordings, a numerator of zero cannot logically be obtained.)

It is a fallacy that a large-magnitude likelihood ratio in one direction or the other equals certainty of individualization or exclusion.

> [S]tatistical evidence, properly interpreted, can be misleading. ... It is the nature of statistical evidence that it can be misleading—it is possible to observe strong evidence supporting $H_2$ over $H_1$ when $H_1$ is true. However, it is also true that strong misleading evidence cannot occur very often. (Ref 28, p. 761)

We cannot expect that evidence (whether statistical or not) will never be misleading. If this were the case we would not need likelihood ratios at all. A likelihood ratio is a probabilistic expression and as such can always be outweighed by other probabilistic evidence (or prior beliefs). Imagine some evidence (e.g., DNA) points towards guilt, but all the other evidence (including any other forensic evidence and any non-forensic evidence such as alibi and eyewitness evidence) points in the opposite direction. The trier of fact should be able to consider whether the latter outweigh the former, and may well conclude that they do even if the likelihood ratio from the former is very large (and even if they restrict consideration to the source level, rather than considering activity-level, contamination, or laboratory-error explanations which could make the source-level likelihood-ratio value of the former moot rather than outweighing it).

## How best to present the results of a forensic analysis

Lennard identifies what we agree is an important question: What is the best way to present the results of a forensic analysis to the trier of fact such that it will be well understood? We think the answer to this question should be sought after first accepting the principle that the trier of fact must be informed as to the logic, assumptions, decisions, and procedures which underlie the result. Despite Lennard's stated concern "that the weight of the evidence we present in court is fully understood and not misinterpreted" (Ref 1, p. 1), we do not believe that his recommendations are compatible with this principle.

## Conclusion

We have argued that, properly explained and understood, the most appropriate way to estimate the strength of forensic evidence is not via an "expert opinion", but rather via a likelihood ratio calculated on the basis of relevant data, quantitative measurements, and statistical models. This does not imply that the former is necessarily useless in all situations, but the latter is more transparent, is easier to replicate, is easier to test under conditions reflecting those of the case under investigation, and is more robust to cognitive bias. Properly explained it makes explicit the question which the strength-of-evidence statement answers and makes explicit the assumptions underlying the procedure for estimating the strength of the evidence. Both of these are important and should be debated before and understood by the trier of fact.